\documentclass[mathleft]{an}
\usepackage{graphicx}
\usepackage{times}
\begin{document}


\title{Rotation curves of luminous spiral galaxies}

\author{I.A. Yegorova,\inst{1}\fnmsep\thanks{Corresponding author:
  \email{iyegorov@eso.org}\newline}
A. Babic\inst{2}, P. Salucci\inst{2}, K. Spekkens\inst{3} \and A. Pizzella\inst{4, 5}
}
\titlerunning{Rotation curves of luminous spiral galaxies}
\authorrunning{I.A. Yegorova, A. Babic, P. Salucci, K. Spekkens \& A. Pizzella}
\institute{
ESO Chile, Alonso de Cordova 3107, Santiago 19001, Chile
\and
SISSA  International School for Advanced Studies, via Bonomea 265, 34136 Trieste, Italy
\and
Department of Physics, Royal Military College of Canada, P.O. Box 17000, Stn Forces, Kingston, Ontario, Canada
\and
Dipartimento di Astronomia, Universit\`{a} di Padova, vicolo dell'Osservatorio 3, I-35122 Padova, Italy
\and
INAF, Osservatorio Astronomico di Padova, Padova, Italy}

\received{}
\accepted{}
\publonline{later}

\keywords{galaxies: kinematics and dynamics -- galaxies: spiral -- galaxies: structure}

\abstract{We have investigated the stellar light distribution  and the rotation curves of high-luminosity spiral galaxies in the
local Universe. The sample contains 30 high-quality extended $H\alpha$ and $HI$ rotation curves. The stellar disk scale-length of these
objects was measured or taken from the literature. We find that in the outermost parts of the stellar disks of these massive objects,
the rotation curves agree with the Universal Rotation Curve (Salucci et al. 2007), however a few rotation curves of the sample show 
a divergence.}

\maketitle

\section{Introduction}

Rotation curves (RCs) of spiral galaxies, if unperturbed, to a very  good
approximation indicate  the galaxy's circular velocity profile $V(R)= (R d\Phi/dR)^{1/2}$, where $\Phi$ is the Newtonian
gravitational potential and  $V^2(R)=V^2_\mathrm{D}(R)+V^2_\mathrm{H}(R)$ is the quadratic sum of the disk and halo
contributions (e.g. Salucci et al. \cite{Salucci-2011}). In combination  with surface photometry, they provide the mass 
distribution in spirals.

It is well known that the distribution of stellar light in a spiral galaxy with characteristic radius
$R_\mathrm{opt} \equiv 3.2 R_\mathrm{D}$, where $R_\mathrm{D}$ is the Freeman exponential thin disk scale-length, does not match 
the distribution of the gravitating matter (Bosma \& van der Kruit \cite{Bosma}; Rubin et al. \cite{Rubin-1};
Persic \& Salucci \cite{Persic+Salucci-90}). To frame this phenomenon it is helpful to define the following reference 
radius $R_\mathrm{opt}$ of the stellar disk: i) it is proportional to the disk scale-length, that represents the stellar distribution 
(Freeman \cite{Freeman}); ii) it encloses 83\% of the total disk mass. Moreover, at $R_\mathrm{opt}$, the {\it disk}  contribution 
$ V_\mathrm{D}(R)$  to the circular velocity is declining rapidly. Assuming, without loss of generality, that 
$V_{D}(R) \propto R^{\nabla_D(R)}$, (where $\nabla$ is the logarithmic slope) in any spiral we have: 
$\nabla_\mathrm{D}(2.2/3.2  \ R_\mathrm{opt})  = 0$, $\nabla_\mathrm{D} (R_\mathrm{opt}) = -0.27$, 
$\nabla_\mathrm{D} (1.2\  R_\mathrm{opt}) \simeq  -0.5$. 

Then at $R \simeq R_\mathrm{opt}$, $V_\mathrm{D}$ rapidly becomes Keplerian. Furthermore, if at $\simeq R_\mathrm{opt}$  the  stellar 
component dominates the total gravitational potential, the resulting RC must decrease with radius in the region of $2.7R_\mathrm{D}$ - 
$3.5 R_\mathrm{D}$ (the region that will be investigated in this paper).  

The RCs of spiral galaxies show a phenomenology that has led to the concept of the {\it Universal Rotation Curve}\\ (URC).
This was  implicit in  Rubin et al. (\cite{Rubin-2}), pioneered by Persic \& Salucci (\cite{Persic+Salucci-2}, hereafter PS91) and set in  Persic, 
Salucci \& Stel (\cite{Persic+Salucci-4}, hereafter PSS). According to this paradigm, the RCs of disk galaxies can be generally 
represented by $V_\mathrm{URC}(R/R_\mathrm{D}; P_i)$, i.e. by a {\it universal} function of normalized  radius, tuned by one or more
galaxy \\ observational quantities $P_i$ (e.g. the B-band luminosity $M_B$ as in PS91, the I-band luminosity $M_I$ as in PSS, or the virial 
mass $M_{vir}$ as in Salucci et al. (\cite{Salucci-1}, hereafter S+07). Remarkably, just one 
global quantity takes into account for more than 90\% of the RC variance. The URC  phenomenology has replaced  the ``flat RC" paradigm 
and switched  the focus  from  the structure of ``a typical galaxy'' to the typical systematics of the  mass  structure of spirals.
PS91, PSS and S+07 universal velocity  functions
$V_\mathrm{URC}^{PS91}(R/R_\mathrm{D}; M_B)$, 
$V_\mathrm{URC}^{PSS}(R/R_\mathrm{D}; M_I)$,  $V_\mathrm{URC}^{S+07}(R/R_\mathrm{D}; M_{vir})$
are  \\ successive improvements of the URC paradigm.
While a complete  assessment of  the role of  minor parameters in the URC  (e.g. bulge, stellar
surface density) has yet to be performed,  the functions $V_\mathrm{URC}$  obtained in the above studies match
the kinematics of individual RCs very well. At any  radius $R$, $V_\mathrm{URC}$ predicts the circular 
velocities of spirals of known  luminosity and disk scale-length within an error that is one order of magnitude 
smaller than i) the radial variations of the RC  in each object,  ii) the  cosmic variance in the RC amplitudes,
at any fixed radius $R/R_\mathrm{D}$.

\begin{figure}
\includegraphics[width=70mm,height=50mm]{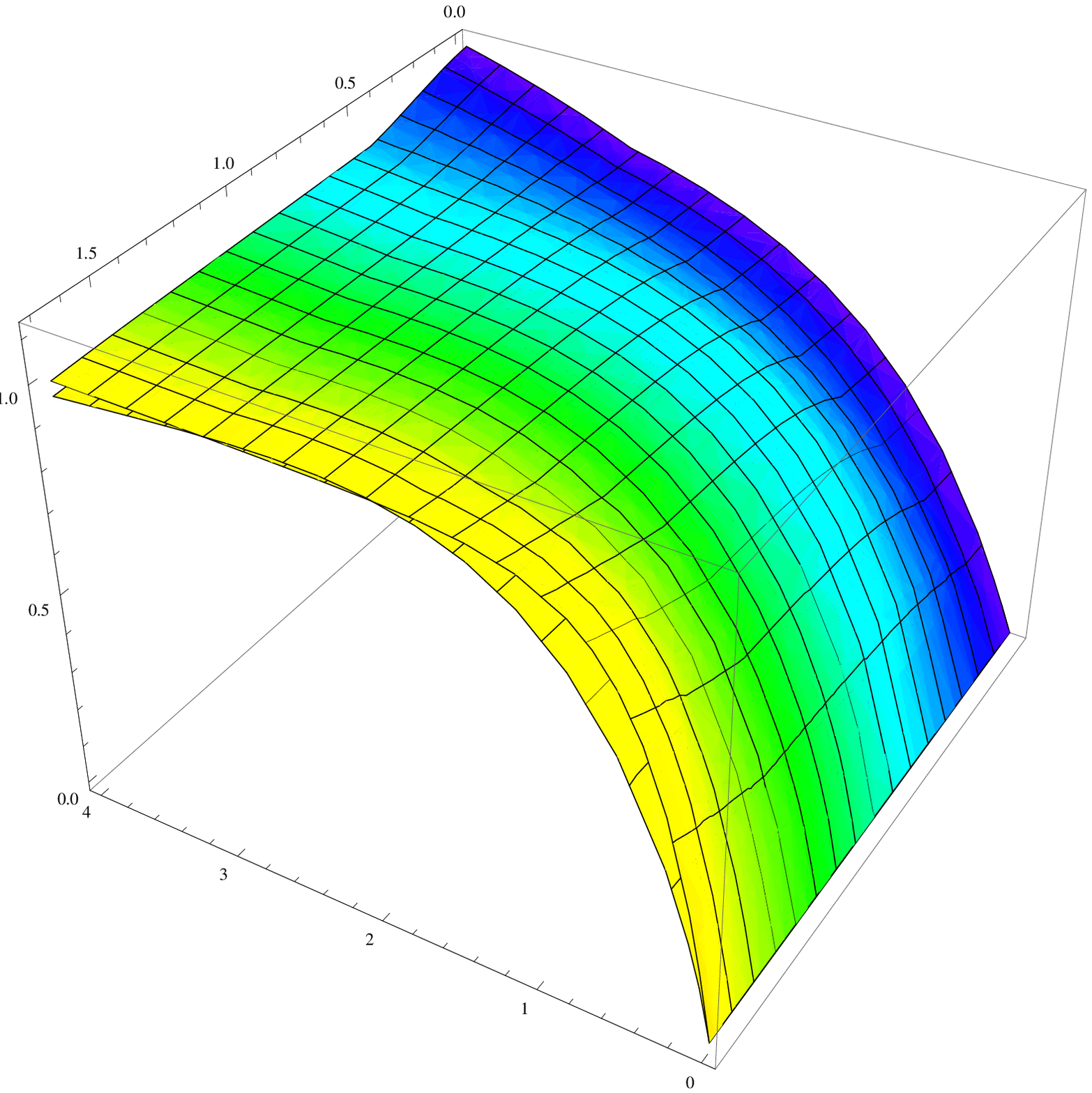}
\includegraphics[width=70mm,height=50mm]{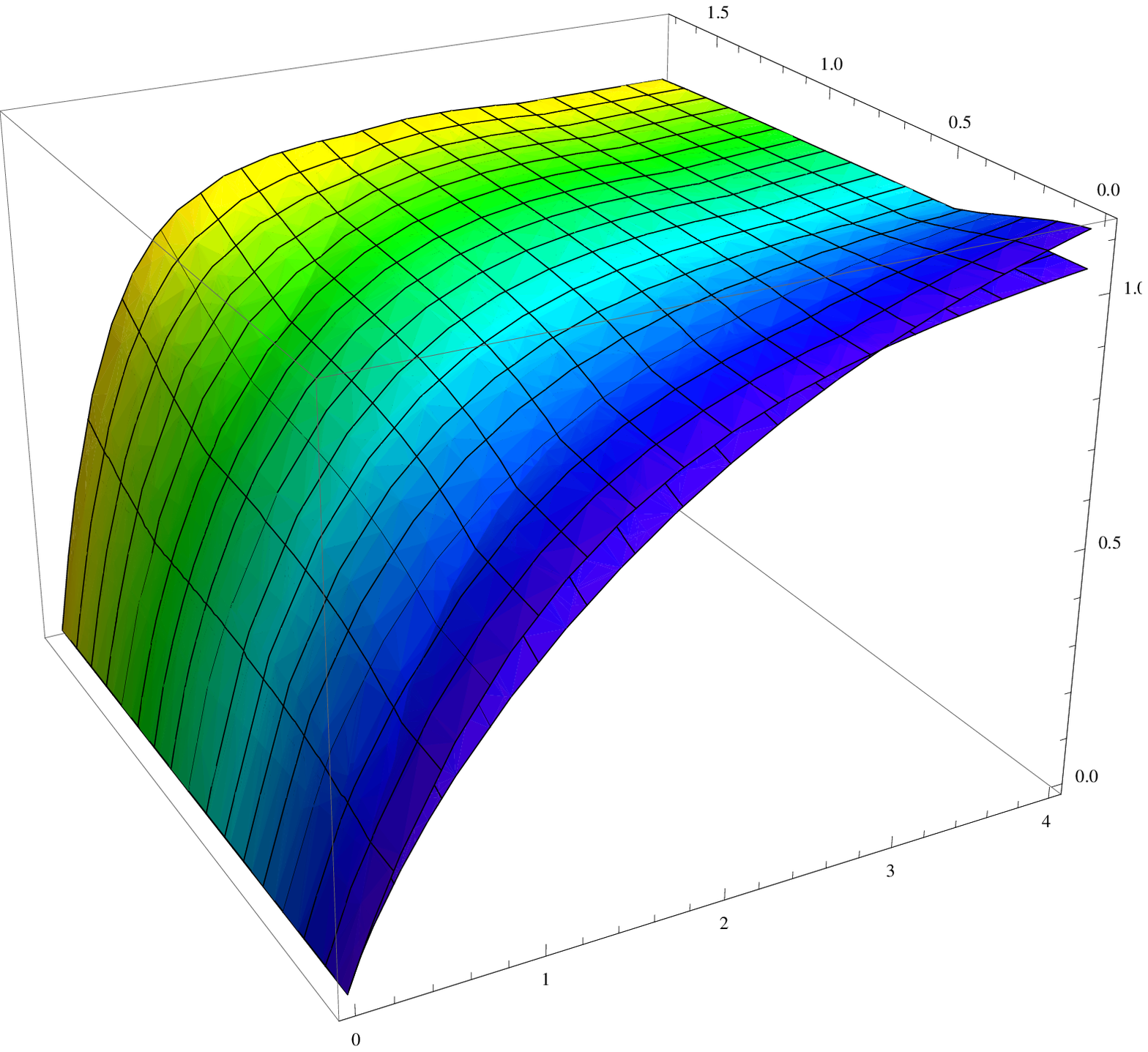}
\caption{Comparison between the S+07 and the C+06 Universal Rotation Curves. 
{\it Upper panel}: the representing 3-D surfaces are plotted together out to 4$R_\mathrm{D}$. We plot 
$V_\mathrm{URC}^\mathrm{S+07}(R/R_\mathrm{D}, M_I)/ V_\mathrm{URC}^\mathrm{S+07}(3.2, M_I)$ 
and $V_\mathrm{URC}^\mathrm{C+06}(R/R_\mathrm{D}, M_I)/ V_\mathrm{URC}^\mathrm{C+06}(3.2, M_I)$ as a function of radius, in units of disk scale-length 
(i.e. of $R/R_\mathrm{D}$) and of “I-magnitudes” (i.e. of $(M_I-19.4)/2.5$). Brighter colors indicate more luminous galaxies. 
{\it Lower panel}: the same surfaces of the upper panel, shown from a different angle.}
\label{URC}
\end{figure}

An important contribution on the  systematic properties of RCs has come  from the work of Catinella et al. 
(\cite{Catinella-et-al-1}, \ hereafter C+06). They studied 2200 RCs of disk galaxies and constructed the 
template RCs for objects of different luminosities. They found  that RC amplitudes and slopes are,  at any radius, 
functions  of the galaxy  luminosity (see Eq. (1), Table 2 and Fig. 1 in C+06). They also found that a generic spiral 
RC is well represented  by  Universal  Templates,  more specifically by a function of radius in disk\\ scale-length units 
and of I-band luminosity:\\ $V_{URC}^{C+06}(R/R_\mathrm{D}, M_I)$. The  S+07 and  C+06 URCs are compared here for the first 
time in Fig. \ref{URC}. They coincide at low and intermediate luminosities.

However, at the high-luminosity end, $M_I<-22$, i.e. for the top 5\% objects in luminosity, from $R>2 R_\mathrm{D}$ onward, there is 
a discrepancy between the C+06 and the S+07 URCs. In this region, in fact, the  profile of the latter slightly declines with radius, while the  
C+06 URC profile, on the contrary, shows a modest but  clear increase with radius.

Due to the high-luminosity cut-off in the spiral galaxy luminosity function, this  issue concerns a small number of objects, which  
however belong to the  important "class" of spirals  with most  extreme properties in term of luminosity/mass  and disk size. The analysis 
of these objects requires extra care with respect to the rest of the spiral population. In fact, these galaxies show  an intrinsically  
``flattish'' rotation curve, so that kinematical effects of the  Grand Design spiral structure or of other not-axisymmetric motions  
are very evident in the RCs as wiggles and oscillations. Combining ``flat''  (and in some cases) non-extended RCs  with large
observational uncertainties or non-axisymmetric velocity components (as done in C+06 and PSS) may introduce biases in the kinematical 
investigations.

Furthermore,  for these objects, we need  a very precise determination of their (flattish) RC slope in order to derive properly 
the underlying  mass  distribution. To show this, and neglecting  for  simplicity of argument the bulge and the HI disk
components,  we recall that the condition of centrifugal equilibrium is

\begin{equation}
\label{velocity}
V^2(R) ~=~ V^2_\mathrm{D}(R)~+~V^2_\mathrm{H}(R),
\end{equation}
where $V(R)$ is the circular velocity we infer from the rotation curve. By defining
$\nabla \equiv ~ ({d{\rm log} V(R) \over d {\rm log} R})$, \\
$\nabla_\mathrm{D}(R) \equiv ~ ({d{\rm log} V_\mathrm{D}(R) \over d {\rm log} R})$, 
and $\nabla_\mathrm{H} (R)$  similarly for the halo component, and the disk fractional contribution to the RC 
as $\beta \equiv { V^2_\mathrm{D}(R) \over V^2(R) }$,  we obtain  the RC-profile equation:

\begin{equation}
\label{nabla}
\nabla(R) ~=~ \beta(R) \, \nabla_\mathrm{D}(R) ~+~ (1-\beta(R)) \,\nabla_\mathrm{H}(R)\ 
\end{equation}
Eq.\,(\ref{nabla})  shows that at a radius $R$ the value of the observed logarithmic slope $\nabla(R)$ is determined 
by the sum of  $\nabla_\mathrm{D}(R)$ and $\nabla_\mathrm{H}(R)$, each one weighted by the corresponding  mass fraction  
inside $R$:  $1-\beta(R)$, and $\beta(R)$. We note that $\nabla_\mathrm{D}(R)$ is known from the photometry and $\nabla_\mathrm{D}(x)$ 
is  equal in all  galaxies for any $x$.

For the sake of simplicity, all quantities, when estimated at $R_\mathrm{opt}$,  drop their subscript, 
e.g. $\nabla \equiv \nabla(R_\mathrm{opt})$. Furthermore, let us recall that, for an exponential thin disk (see Freeman \cite{Freeman}): 
$\nabla_\mathrm{D}(2.2 R_\mathrm{D})=0$, $\nabla=-0.27$, \\ and $V_\mathrm{D}(2.2 R_\mathrm{D})=1.28 V_\mathrm{D}(3.2 R_\mathrm{D})$, 
where $3.2 R_\mathrm{D} = R_\mathrm{opt}$. Supported by observations and successful mass modeling we can assume that, between $2.2 R_\mathrm{D}$ 
(the radius where \\ $dV_\mathrm{D}/dR=0$), and $R_\mathrm{opt}$, both \hspace*{-0.15cm} $V(R)$ and $V_\mathrm{H}(R)$ are roughly linear (e.g. 
in the region 
considered $V(R)\simeq  V_\mathrm{opt} (1+(R/R_\mathrm{opt}-1)\nabla)$. Then, Eq.\,(\ref{nabla}), evaluated at $2.2 R_\mathrm{D}$ 
and at $R_\mathrm{opt}$ leads to the following system of equations

\begin{equation}
\label{}
-0.27 \beta +\nabla_\mathrm{H}-\beta \nabla_\mathrm{H}=\nabla
\end{equation}

\begin{equation}
\label{}
-1.71 \beta \nabla_\mathrm{H} + 2.2 \frac{\nabla} {3.2- \nabla} \nabla_\mathrm{H} =  2.2 \frac{\nabla} {3.2- \nabla}
\end{equation}

\begin{figure}
\includegraphics[width=75mm,height=50mm]{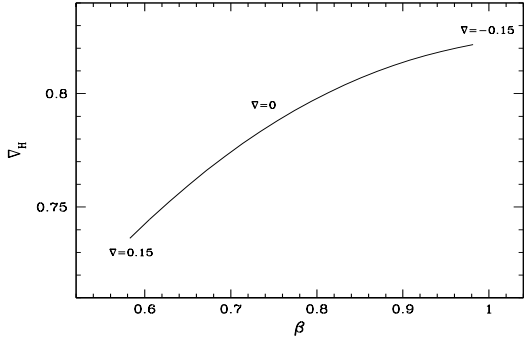}
\caption{$\beta$ and $\nabla_\mathrm{H}$  as a function of $\nabla$.}
\label{as}
\end{figure}

The  value of $\nabla(R)$  ranges in spirals between $\simeq -0.3 $ (corresponding to a negligible amount of dark matter (DM) inside 
$R_\mathrm{opt}$), and  $\simeq 1$ (corresponding to a negligible amount of luminous matter (LM) inside $R_\mathrm{opt}$).  
More specifically, the population of high luminosity spirals (under study here) shows

\begin{equation}
\label{}
-0.15 \leq \nabla \leq 0.15,
\end{equation}
where the above  range  includes the values from the  individual RCs of our sample, and  those derived from  the C+06, and S+07 URCs. 
The solution of  Eq. (3, 4) shown in Fig. \ref{as} is a relationship between $\nabla_\mathrm{H}$ and $\beta$, with  $\nabla$ 
the known parameter. When $\nabla\simeq 0$ a precision of $\delta \nabla < 0.04$ is required to perform 
a proper mass modeling, since larger errors produce unacceptable uncertainties.
In fact, from Fig. \ref{as}  we realize that for the value of $\nabla=0.12$, consistent with that of the  high luminosity 
end of  C+06 URC, we have  $\nabla_\mathrm{H}=0.72$  and $\beta=0.58$, values that are still reproducible (though with some difficulty) 
by a NFW halo + Freeman disk mass model (NFWD) (Salucci \& Persic \cite{Persic+Salucci-99}),  while, for the value of  $\nabla=-0.12$, 
consistent with that of the high luminosity end  of the S+07 URC  we have: $\nabla_\mathrm{H}=0.9$ and $\beta=0.95$, values that cannot 
be accounted by any NFWD mass model.

In building the URC at the Highest Luminosity End \\ (HLE), one should use only reliable RCs. However only 9  HLE RCs in  
PSS, and not many more in  C+06 are of  such quality, e.g. can  be used for individual mass modeling.

\hspace*{-0.2cm} By inspecting the HLE  RCs  in the region  between $2.7R_\mathrm{D}$ and $3.5 R_\mathrm{D}$
we realize that they  flatten and then, at least  in some cases, start to decline. Over a limited region, centered on $\simeq 3 R_\mathrm{D}$ and 
$(0.5-0.7) \ R_\mathrm{D}$  wide, $\nabla(R)$ passes from positive values $\sim 0.15$,  to zero and, in some cases, to also negative values down to 
$\sim \ -0.15$. This RC behaviour, unique in the spiral RC phenomenology, requires that for HLE the disk scale-length $R_\mathrm{D}$ be known to good 
precision. In fact, if we underestimate its value by more than 15\%, we may obtain, for a negative $\nabla$ an incorrect positive estimate (as a 
consequence of measuring at $2.7 R_\mathrm{D}$ rather than at $R_\mathrm{opt} \equiv 3.2 R_\mathrm{D}$).

Considering  also works published  after 2007, the aims of this paper are 1) to take or build from raw literature data all
available/possible  smoothed high quality RCs of high-luminosity spirals; 2) to measure or take from literature their disk
scale-length and then, for these objects; 3) to investigate the individual RC profiles; 
and 4) explore the phenomenology of HLE  RCs.

\section{Data selection}

We select high-quality RCs of HLE late-type spirals (Sb or later) that satisfy the following criteria:

\begin{enumerate}
\item the  objects  are at the high end of the spiral luminosity/velocity range: $220\  \mathrm{km/s} < V(R_\mathrm{opt}) < 335\  \mathrm{km/s}\ $ (corresponding to $-22.4 < M_B < -20.6$);
\item the inclination  $i$ of the galaxies  is $57^\circ <$ \textit{i} $< 90^\circ$, ensuring sufficiently small error in the normalization of the RC;
\item the RCs have a good spatial resolution: the distance between two  consecutive velocity measurements must be $\leq 0.5$ $R_\mathrm{D}$;
\item the RCs are sufficiently well-sampled: at least 4 independent RC points lie in the region of interest: $2 < R/R_\mathrm{D} < 4$;
\item the RCs are  unperturbed, ensuring  that they reflect the gravitational potential. Moreover, they do not exhibit features indicative of prominent spiral structure nor \\ warps;
\item the RCs have small observational uncertainties in their amplitudes and derivatives: $\delta  V/ V < 0.04  $, $\delta  \nabla <0.05 $, in order to  ensure a  sufficient precision in the determination of $\nabla(R)$.

\end{enumerate}

These criteria are just those necessary to: a) identify the objects that we want to study, and b) ensure that we obtain reliable RCs. 
No result of this paper depends critically on this (very reasonable) selection process.  

\section{The disk scale-length}

A reliable estimate of the disk scale-length $R_\mathrm{D}$ is fundamental in the sample selection.
Looking at the samples of potential RCs, we must notice that in Vogt et al. (\cite{Vogt-et-al}) and C+06 samples
the available $R_\mathrm{D}$ were derived from I-band imaging observations of  Haynes et al. (\cite{haynes}). The
authors fitted only the outer parts of the  disk, since their aim was to extrapolate the light profile beyond
the outer edge  and obtain the total magnitude to use in establishing the Tully-Fisher relation.
For the objects in the  Courteau (\cite{Courteau}) sample, the available $R_\mathrm{D}$ were obtained from
exponential fits to the radial  r-band  surface brightness   profile,   performed over the region $50-90\%$
of the radius at 26~$r\,\mathrm{mag\,arcsec}^{-2}$, again over the  outer disk region.
These  determinations of $R_\mathrm{D}$ based on  outer surface brightness data is certainly biased and often incorrect.
The proper analysis requires that the latter  be decomposed in the separate contributions of a bulge and
a disk and that, obviously,  the whole data are taken into account. Among several drawbacks that  the
former procedure can have, we draw the attention to the following a) due to extinction, especially in  inclined galaxies,
surface brightness profile will steepen towards the outer edge, and the scale-length measured in the outer disk will
be smaller than the scale-length measured  over the whole disk; b) to neglect the bulge can affect
the values of the disk scale-length in an unpredictable way.

\begin{figure}[h!]
\vspace{-15 mm}
\includegraphics[width=60mm,height=80mm, angle=270]{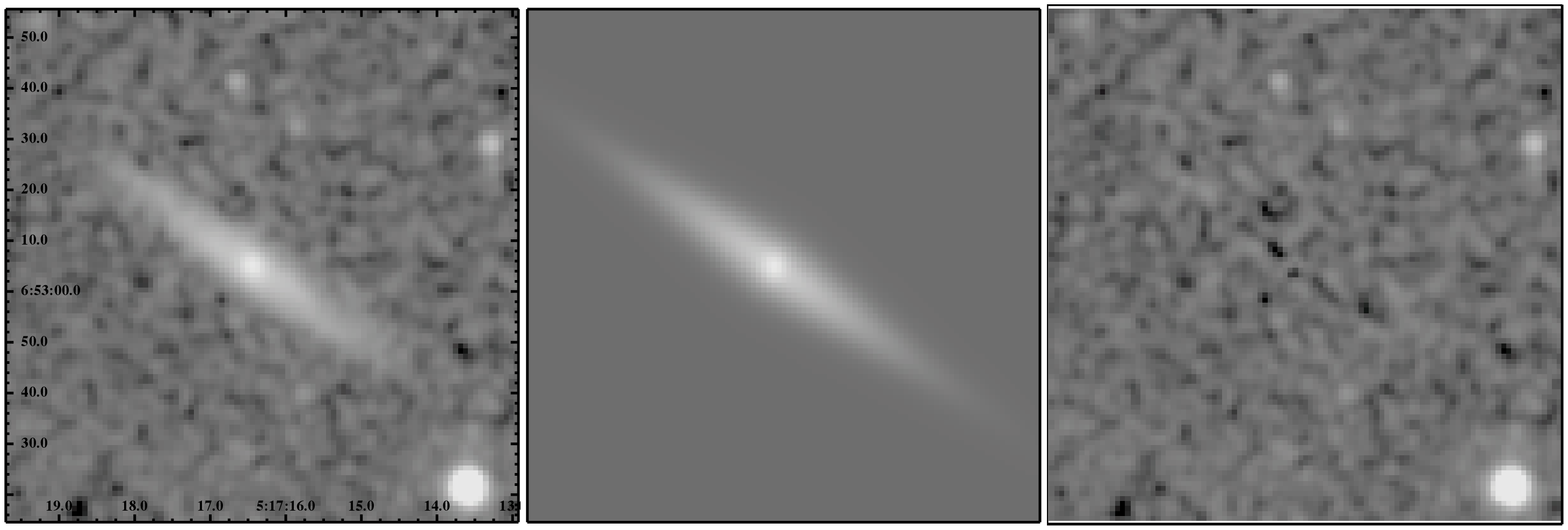}
\vspace{-10 mm}
\caption{Two-dimensional 2MASS $K_\mathrm{s}$-band image decomposition for UGC~3279. From left
to right: the original image, the model image, and the residual image.}
\label{galfit-1}
\end{figure}

\begin{figure}[h!]
\includegraphics[width=70mm,height=70mm, angle=270]{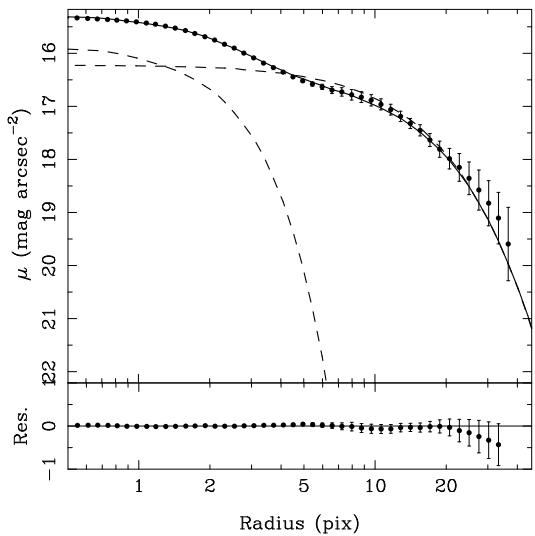}
\caption{Surface brightness profile of UGC 3279. The image
profile is marked by filled circles, the complete
model (obtained from 2-D image decomposition)
with a solid line, and the disk-bulge components of the model
are marked with dashed lines. The model-data residuals are also
shown in the bottom panel.}
\label{galfit-2}
\end{figure}

Thus, for 15  candidate objects in  the above samples we opted to re-derive the disk
scale-lengths from  available near-infrared images, by using the two-dimensional disk-bulge decomposition code GALFIT
(Peng et al. \cite{galfit}). All the images are taken from the $K_s$-band 2MASS All Sky Release Atlas.
They were photometrically calibrated, and the typical FWHM for the
point-spread function (PSF) is $\sim 3\arcsec$. Two components were used in the
fits: an exponential disk and a S\'{e}rsic bulge or a PSF for unresolved central
sources. For a number of objects, the bulge half-light radius is comparable
to the PSF FWHM, and the corresponding $r_e$ estimates should be considered unreliable.
In Fig. \ref{galfit-1} we show an example of a successful two dimensional image decomposition,
and in Fig. \ref{galfit-2} we show the corresponding one-dimensional surface brightness
profiles. The derived disk scale-lengths $R_\mathrm{D}$ are reported in the Table~\ref{t1} and  marked with an asterisk.

The new disk scale-lengths values tend to differ by 10\% to 50\%  from those in C+06, Vogt et al. (\cite{Vogt-et-al}) 
and Courteau (\cite{Courteau}). In 5 cases the new 
(larger) scale-lengths  made the available  RC not extended enough to be included in the  sample (see Table~\ref{t2}, we report 
these values since they could be useful in future studies). Of course, we also found in the literature properly determined disk 
scale-lengths, that we have adopted in our studies (see Table~\ref{t1}).

\section{The Sample}

At the end of the selection procedure we have a sample of 30 {\em high-quality} RCs of HLE spirals (see Table~\ref{t1}). 
In detail we have: 1) H$\alpha$ RCs: 2 from Courteau (\cite{Courteau-2}), 9 from Vogt et al. (\cite{Vogt-et-al}), 
5 from C+06, 1 from Blais-Ouellette (\cite{Blais-Ouellette}), and 8 from PSS. 2) HI RCs\footnote{The RCs from 
Spekkens et al.\ (\cite{Spekkens}) combine both H$\alpha$ and HI data. But since the kinematics in the region 
of interest ($R>R_\mathrm{D}$) are covered by the HI component, we consider them as HI RCs for the purposes 
of this study.}: 2  from Spekkens et al. (\cite{Spekkens}), 2 from Kregel \& van der Kruit (\cite{Kregel1}); 
Kregel, van der Kruit \& Freeman (\cite{Kregel2}) and 1 from Corbelli et al. (\cite{Corbelli}).

Let us stress  that a number of published RCs (including also some in PSS and  in C+06) did not enter the present 
sample since they are not 1) extended, or 2) smooth, 3) symmetric, or  4) of sufficient spatial resolution, or 
5) with small velocity  r.m.s. We believe that  these requirements are essential to investigate $\nabla$ in HLE spirals.

This is the largest sample  of  high-luminosity spirals ever studied in which every object
has suitable kinematics and photometry.  Unfortunately,
the number of objects is still limited and the results obtained must be considered as indications.
Let us stress  that, at this point in time, this is the best we can do: there are additional
high luminosity  objects with kinematical data in the literature, but they do not satisfy the 
criteria for inclusion in our sample (e.g. UGC 8707 (Courteau \cite{Courteau-2}) and UGC 1901, AGC 250022 (C+06)).

\section {The Rotation Curves}

The 30 raw  RCs have been  binned as in Yegorova \& Salucci (\cite{Yegorova}),  i.e. smoothed with respect to
non-axisymmetric motions and observational errors. We adopted radial bins sizes of $0.2 R_\mathrm{D}$
and $0.6 R_\mathrm{D}$ for $H\alpha$ and $HI$ RCs, respectively. We plot the  RCs in the Appendix.

\begin{table*}
\begin{center}
\caption{Data: the absolute B magnitude are calculated using the apparent B-magnitude from HyperLeda database.
Distance: is taken from the NED NASA database. Reference: Vogt et al. (\cite{Vogt-et-al}) - 1, Courteau (\cite{Courteau-2}) - 2,
C+06 - 3, Kregel \& van der Kruit (\cite{Kregel1}); Kregel, van der Kruit \& Freeman (\cite{Kregel2}) - 4, Blais-Ouellette et al. 
(\cite{Blais-Ouellette}) - 5, Spekkens et al. (\cite{Spekkens}) - 6, PSS - 7, Corbelli et al. (\cite{Corbelli}) - 8.} \label{t1}
\label{tlab}
\begin{tabular}{cccccccc}\hline
Name    &$M_B$& Inc &$\nabla$&Distance&$R_\mathrm{D}$ & Data &Ref\\
        &  (mag) & ($^\circ$) && (Mpc) &   (kpc) & &\\
\hline
UGC944    &-21.37& 81 &-0.15&68.5&2.9&$H\alpha$&1\\
UGC1076   &-22.08& 72 &-0.05&179&5.5*&$H\alpha$&3\\
UGC1094   &-20.97& 80 &-0.13&61.6&3.4&$H\alpha$&1\\
UGC3279   &-21.92&  82 &0.21&114&5.4*&$H\alpha$&1\\
UGC4895   &-20.82& 70  &0.01&96.1 &3.5 &$H\alpha$&1\\
UGC4941   &-22.37& 83 &0.03&83.2&3.0&$H\alpha$&1\\
UGC8140   &-21.0& 78&-0.02 &97.4&4.8&$H\alpha$&1\\
UGC8220   &-21.14& 86&-0.04&97.7 &4.5*&$H\alpha$&1\\
UGC9805   &-21.13&  68 &-0.02&47.3&3.3*&$H\alpha$&3\\
UGC10692  &-21.47& 77   &-0.25&130&8.6 &$H\alpha$&3\\
UGC10815  &-20.77& 78&0.01 &54.9 &4.0*&$H\alpha$&2\\
UGC10981  &-21.84& 66  &-0.16 &151&5.1*&$H\alpha$&1\\
UGC11455  &-21.81& 84 &0.1 &76.8 &2.2&$HI$&6\\
UGC11527  &-20.63&  78 & -0.08&77.2 &3.5*&$H\alpha$&3\\
UGC12200  &-21.49& 57 &0.01&136&3.8*&$H\alpha$&2\\
UGC12678  &-21.28& 90  &-0.03 &127&5.2*&$H\alpha$&1\\
AGC241056 &-22.08& 80&-0.02 &293&8.0* &$H\alpha$&3\\
NGC1247   &-21.81& 90 &-0.12&53.4& 4.8 &$H\alpha$&7\\
NGC5170   &-20.89& 90 &-0.11&19.1&6.5 &$HI$&4\\
NGC5985   &-21.4& 58 &-0.08&36.7&5.3 &$H\alpha$&5\\
NGC7300   &-21.48&72&-0.04& 68.1& 4.4 &$H\alpha$  &7 \\
IC2974  &-21.8 &86  &-0.13&76.9& 5.5 &$H\alpha$ &7\\
ESO141-G20&-21.98 & 90  &-0.07&59.9& 4.5 & $H\alpha$ &7\\
ESO141-G34&-21.65 & 90 &-0.02 &59.1&  5.3 & $H\alpha$ &7 \\
ESO240-G11&-21.41& 90& -0.01&38.3&9.1 &$HI$&4\\
ESO350-G23&-21.56 & 77& 0.05 &21.4& 4.3 & $H\alpha$ & 7\\
ESO374-G27&-21.60 &62  &-0.1&122& 5.5 &$H\alpha$  &7 \\
ESO563-G21&-22.0& 83 &0.04&59.9&2.0 &$HI$&6\\
ESO601-G9&-22.37&90 &0.03&37& 5.4 & $H\alpha$ &7 \\
M31&-21.71&77 &0.04&-&4.5& $HI$ &8 \\

\hline
\end{tabular}
\end{center}
\end{table*}

We do not consider here the innermost regions of the RCs, some dominated by a bulge, and we plot RCs only for
$R > R_\mathrm{D}$. In a small number of cases, the contribution 
from the central bulge to the dynamics of the galaxy is not negligible. At this radius this does not affect 
the present results, in that, at larger radii these RCs are very similar to the others. However, an  analysis 
like this one for spirals with more significant bulges  is certainly worthy of future attention. In the website 
indicated in  Salucci et al. (\cite{Salucci-2011}) the whole 30 RC data are available for download.

We focus  on the behavior of the  RCs near $R_\mathrm{opt}$. 
We compute the logarithmic slope $\nabla$ by fitting  linearly the RC in the neighborhood of $ R_\mathrm{opt}$.
The logarithmic slope $\nabla$  is estimated within  an uncertainty  of about  0.05.  We therefore,  define  a (slowly) 
declining RC when $\nabla(R_\mathrm{opt}) \leq   -0.05$, a flat RC when  $-0.05 \leq \nabla(R_\mathrm{opt}) \leq 0.05$ 
and a  rising RC   when $0.05 \leq \nabla(R_\mathrm{opt})$.

We find that at $R_\mathrm{opt}$   40\% of the RCs are declining, 50\% flat and 10\% slowly
rising. However, there are cases when a ``rising''  RC  (at $R_\mathrm{opt}$), shows a decline farther out (e.g. ESO350-G23).

It is evident that adopting an incorrect  value for $R_\mathrm{opt} \simeq 3.2 R_\mathrm{D}$ would bias the analysis.
When the adopted $R_\mathrm{opt}$ is smaller than the actual one, the error  leads  to  a wrong positive estimate of $\nabla(R_\mathrm{opt})$
(since $V(R)$ in the neighborhood  of $R_\mathrm{opt}$ turns over and begins to decline). When instead the adopted $R_\mathrm{opt}$ 
is larger than the actual one, the resulting value of $\nabla(R_\mathrm{opt})$ is (roughly) correct since the HLE RCs for $R> R_\mathrm{opt}$
are  already converged to a linear/flat profile). Thus, adopting the C+06 disk scale-lengths leads to an increase in fraction 
of (apparently) rising RC's. This, with the fact that some rising RC do exist, explains the noted discrepancy between the S+07 and C+06.

\begin{figure}
\includegraphics[width=82mm,height=57mm]{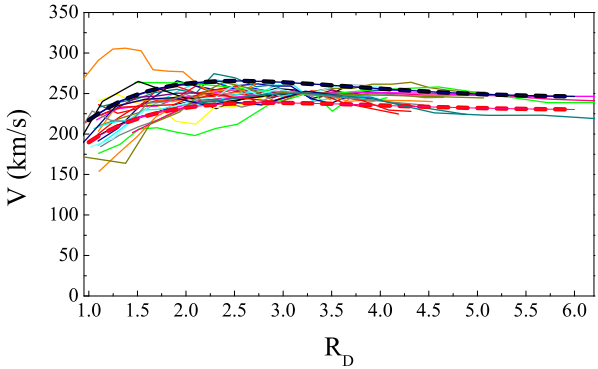}
\caption{The RCs of the sample normalized to the same $V(R_\mathrm{opt})$. The dashed red and black lines indicate the URC predictions for
$V(R_\mathrm{opt})= 220 \   \mathrm{km/s}$ and $V(R_\mathrm{opt})= 350 \   \mathrm{km/s} $ correspondingly.}
\label{color}
\end{figure}

We plot the  RCs of our sample all together in Fig. \ref{color}. They are normalized by setting
$V(R_\mathrm{opt}) = 250\ \mathrm{km/s}$ in  each  galaxy, with the normalized velocity curve defined
as $250\  V(R)/V(R_\mathrm{opt}) \ \mathrm{km/s} $. This  specific  value was adopted for visual clarity 
and comparison of RCs with different \\ $V(R_\mathrm{opt})$. 

We overplot (with red and black dashed lines, see Fig. \ref{color})
the URC {\it profile}  corresponding to $V(R_\mathrm{opt})= 220 \   \mathrm{km/s} $  and $V(R_\mathrm{opt})= 350 \   \mathrm{km/s}$ 
as given by S+07. We note that the URC, in general, well represents the individual profiles, but there are notable exceptions in which  
we detect a rising RCs (e.g. UGC~3279, UGC~11455).

\section{Discussion}

We have investigated the RCs of high-luminosity spirals, a class of objects whose kinematical properties had not yet 
been thoroughly examined so far. The sample consists of 30 high-quality, extended $H\alpha$ and $HI$ rotation 
curves with $220 \  \mathrm{km/s} \leq  V(R_\mathrm{opt}) \leq 335\ \mathrm{km/s}$ that represent, in this paper, 
the HLE spirals. We found that, {\it to a first approximation}, HLE have a mass distribution that follows the same phenomenology 
of spirals of smaller mass.  

We proved this by examining and studying each RC individually. We determined that the discrepancy {\it at the high-luminosity end} between the S+07 URC  
and  the  C+06 URC originates from differences in the way that scale-lengths are estimated in these two studies. Nevertheless, in  partial agreement 
with C+06, a presumably {\it small} fraction  of HLE spirals with gently rising RC out to $R_\mathrm{opt}$ seems to exist. The individual 
study of these objects  will be very important. 

At $R_\mathrm{opt}$, the radius  where the velocity profile of the luminous components starts to decrease, the RCs of our sample divide between 
declining, flat, and  (a few) mildly rising RCs.

\begin{figure}
\includegraphics[width=72mm,height=45mm]{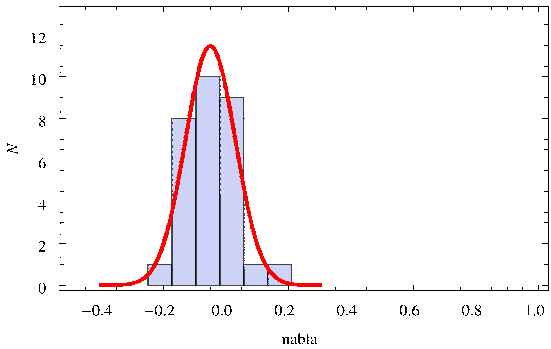}
\caption{Histogram of the values of $\nabla$'s for the objects of our sample, compared with the predictions of the S+07 URC (red solid line).}
\label{nablavar}
\end{figure}

Let us compare the distribution of $\nabla$'s as found  in the present sample with the one that comes from the S+07  URC
predictions, i.e. a Gaussian centered at $\nabla=-0.05$ with width of $0.1$ (see Fig.~\ref{nablavar}). They are in a very good agreement considering that,  
in principle, $\nabla$ can take {\it any} value between $-0.5$ and $1$. In general, HLE RCs seem to be in agreement with the S+07 URC.

Most of the RCs of our sample exhibit a clear decline at some radius $R>2 \ R_\mathrm{D}$ indicating that  
the stellar disk contributes to the mass budget in this regime. However, in all cases $ \nabla > \nabla_\mathrm{D}\simeq -0.27 $, which implies 
the presence of a dark matter halo. 

Although to derive the properties of the dark and luminous matter in HLE requires their individual  mass modeling, some important feature can  still be  
obtained, also  from a simpler RC analysis.  By means of the equations shown in the Introduction  we obtain that at $R_\mathrm{opt}$: 1) $0.7 \leq \beta \leq 0.9$ i.e. 
high luminosity objects are luminous matter dominated, and 2) without any loss of generality, the halo density profile can be written as
$\rho_\mathrm{H}(R)\propto R^{2(\nabla_\mathrm{H} -2)}$, therefore the density slope emerges being very shallow $0.77 \leq  \nabla_\mathrm{H} \leq 0.82 $.

As in spirals of lower luminosities (see PSS), we do not see any ``cosmic conspiracy'', i.e. any   fine-tuning among the values of  $P_i$ describing the  DM/LM distributions 
(the halo core radius and the central density, and the disk mass) that ``creates'' the observed  ``flattish'' RC profile. On the contrary, the RCs of high luminosity spirals 
show quite large ranges in the values of the $P_i$,  that,  instead,  turn out to be very correlated. The uncertainty ranges are relevant, they indicate the existence of a real 
variance of the mass distribution of HLE  (as represented by our sample).

The framework of a NFWD requires  $\nabla(R) >  0.05 \simeq 0.1$ from $1 \ R_\mathrm{D}$ to $6 \ R_\mathrm{D}$ 
(see Salucci \& Persic \cite{Persic+Salucci-99}). This prediction is  not fulfilled by the RC profiles  of our spirals: many of them,  {\it decline}  i.e. have   
$\nabla(R) <0$, over $>1$ disk scale-length (see Table \ref{t1} and Fig.~\ref{nablavar}). Therefore, also  in  high luminosity galaxies, a disagreement between 
{\it naive}  $\Lambda$CDM  predictions  and the  actual data does clearly emerge.
 
We conclude this paper  by indicating  possible causes  of  the  (moderate)  cosmic variance of the RC profiles of HLE:
effects of the environment, the presence of an active galactic nucleus,  the stellar and central black hole feedback  are possibilities that need to  be investigated  
with a much larger sample of objects.

\begin{table}[h!]
\begin{center}
\caption{Measured disk scale-lengths for galaxies not in final sample.} \label{t2}
\label{tlab}
\begin{tabular}{cccc}\hline
Name    &$M_B$& $R_\mathrm{D}$ \\
        &  (mag) &   (kpc)  \\
\hline
UGC562&-22.21   &6.8*   \\
UGC8017&-20.89  &3.1* \\
AGC24797&-21.57 &4.7* \\
AGC211561&-21.51 &5.5* \\
AGC320581&-21.1  &4.6*  \\

\hline
\end{tabular}
\end{center}
\end{table}

\acknowledgements
We thank the anonymous referee for valuable comments that improved the paper. A.P. acknowledge funding through grants CPDA089220/08 by Padua University and ASI-INAF I/009/10/0.

\appendix

\section{RCs of the sample}

\begin{figure*}
\includegraphics[width=170mm,height=230mm]{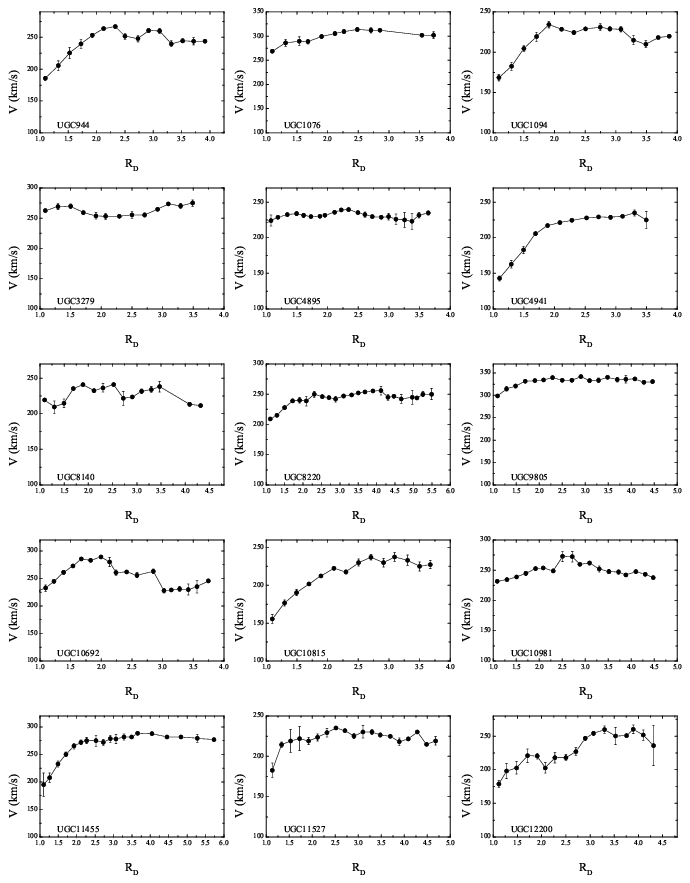}
\caption{RCs from our sample.}
\label{RC1}
\end{figure*}

\begin{figure*}
\includegraphics[width=170mm,height=230mm]{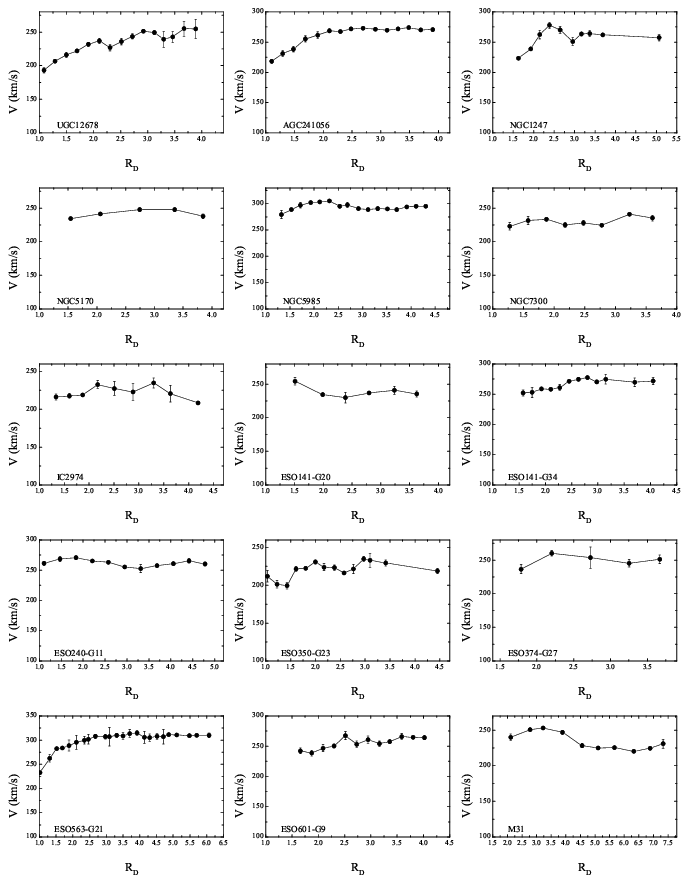}
\caption{RCs from our sample.}
\label{RC2}
\end{figure*}

\end{document}